\documentclass[journal]{IEEEtran}
\usepackage{cite}

\usepackage{graphicx}
\graphicspath{{./images/}}  
\usepackage{array,booktabs}

\ifCLASSOPTIONcompsoc
\usepackage[caption=false,font=normalsize,labelfont=sf,textfont=sf]{subfig}
\else
\usepackage[caption=false,font=footnotesize]{subfig}
\fi
\usepackage{dblfloatfix}
\usepackage{amsmath}

\begin{document}
	%
	\title{Linear Delay-cell Design for Low-energy Delay Multiplication and Accumulation}
	%
	%
	%
	
	\author{Aditya~Shukla,~\IEEEmembership{Student Member,~IEEE}
		\thanks{The author is with the Department
			of Electrical Engineering and Computer Science, University of Michigan, Ann Arbor,
			MI - 48104, USA (e-mail: aditshuk@umich.edu).}
	}

	\maketitle
	
	\begin{abstract}
		A practical deep neural network's (DNN) evaluation involves thousands of multiply-and-accumulate (MAC) operations. To extend DNN's superior inference capabilities to energy constrained devices, architectures and circuits that minimize energy-per-MAC must be developed.
		In this respect, analog \emph{delay}-based MAC is advantageous due to reasons both extrinsic and intrinsic to the MAC implementation $-$ (1) lower fixed-point precision requirement for a DNN's evaluation, (2) better dynamic range than charge-based accumulation, for smaller technology nodes, and (3) simpler analog-digital interfacing.
		Implementing DNNs using delay-based MAC requires mixed-signal delay multipliers that accept digitally stored weights and analog voltages as arguments.
		To this end, a novel, linearly tune-able delay-cell is proposed, wherein, the delay is realized using an inverted MOS capacitor's ($C^*$) steady discharge from a linearly input-voltage dependent initial charge. The cell is analytically modeled, constraints for its functional validity are determined, and jitter-models are developed.
		Multiple cells with scaled delays, corresponding to each bit of the digital argument, must be cascaded to form the multiplier.
		To realize such bit-wise delay-scaling of the cells, a biasing circuit is proposed that generates sub-threshold gate-voltages to scale $C^*$'s discharging rate, and thus area-expensive transistor width-scaling is avoided.
		For 130nm CMOS technology, the theoretical constraints and limits on jitter are used to find the optimal design-point and quantify the jitter versus bits-per-multiplier trade-off. Schematic-level simulations show a worst-case energy-consumption close to the state-of-art, and thus, feasibility of the cell.
	\end{abstract}
	

	%
	\IEEEpeerreviewmaketitle

	\section{Introduction}
	\IEEEPARstart{R}{ecent} advances in machine learning algorithms and, particularly, deep neural networks (DNNs), have equipped portable computing devices with human-like inferring, classifying and planning capabilities.
	Enormous sizes of these networks, with number of operations per evaluation often running into millions, make remote computing servers indispensable.  
	Reliance on servers increases inference latency, communication energy, risk of privacy loss, traffic, and needs a perpetual connection to the server. Some of these metrics are critical in applications like self-driven cars, that cannot afford delays while making decisions.
	Delocalizing computational effort for evaluating ML model, away from server and towards the leaf nodes, requires ML-specific energy-efficient computing architectures \cite{Xu2018ScalingNetworks}.
	Many such architectures have been proposed to greatly accelerate the training and inference speed of DNNs \cite{Jouppi2017In-DatacenterUnit,Fowers2018AAI,Sze2017EfficientSurvey}, but much work is needed to efficiently run these networks under severe energy restrictions many portable devices operate under.
	\par
	The computing-energy's problem \cite{Horowitz2014ComputingsIt} is tackled by: (1) using simpler data-types: these algorithms do not require a large precision and continue to provide similar accuracy with simpler data-types and restricted widths \cite{Judd2016Stripes:Computing,Reagen2016Minerva:Accelerators,Rastegari2016XNOR-Net:Networks,Sharify2017Loom:Networks,Courbariaux2014Binarized1,MoonsBinarEyeCMOS} (2) minimizing data-transfer: the number of data-fetches shoots up for human-level, large-scale applications of these algorithms causing significant non-compute (latent) energy losses \cite{Jouppi2017In-DatacenterUnit,Sze2017EfficientSurvey,Chen2014DianNao}.
	\par 
	Relative robustness of DNNs to precision-loss, together with a limitation on energy, motivates the use of analog computing systems, wherein, the loss of information due to noise and process-variability can effectively be modeled as the loss in precision. To maintain the energy-efficiency without an excessive (counter-productive) precision-loss, these systems constitute both analog and digital computing units. The computational roles are distributed such that the multiply-and-accumulate (MAC) operations, which form the bulk of a DNN's evaluation, are executed in an analog domain, while other operations (e.g. control-flow, data-communication and storage) are done using binary voltages. Superposable electrical variables like charge \cite{Kang2018AArray,Lee2017AnalysisComputing} and current \cite{Li2017ResistiveMethodology,Skrzyniarz201624.3CMOS,Wang2017ASensors} physically represent partial sums of a MAC, with a capacitor as a an accumulator to store the sum of physical variables.
	
	\par 
	Recently, time was proposed as an accumulation variable, as it is better than charge and current in following regards: (1) time-to-voltage/digital converters (TDC, and vice versa – DTC) are more area and power-efficient than voltage-based converters \cite{Miyashita2014AnProcessing,Li2009Delay-Line-BasedConverters}. For instance, both DTC and TDC can be realized out of clocked counters, while voltage ADC/DACs require area and energy-expensive operational amplifiers; (2) while the noise-floor is relatively constant, the supply voltage, $V_{DD}$, drops with technology nodes. Thus, the dynamic-range of accumulation of voltage, current or charge gets increasingly limited; (3) the transition frequency of the FETs, which dictates the temporal resolution of a TDC, increases with tech. nodes.
	\par
	Within the purview of time-based accumulation, pulse-width \cite{Sayal201914.4Computing,Sayal2020AComputing} and pulse-delay \cite{Miyashita2014AnProcessing,Miyashita2017AProcessing,Gopal2018ACMOS} are the two modulation schemes that have been demonstrated on-chip. Of these, pulse- (or, event) delay is more promising for MAC applications due to (1) free addition/subtraction in case of delay, and (2) requirement of peripheral pulse re-routing circuitry requirements in the prior. 
	\par 
	A practical delay-MAC must meet following specifications: firstly, it must accept mixed-signal arguments $-$ one analog while other digital, for locally stored weights; secondly, it should posses linear voltage-delay (transfer) characteristics to accept externally sensed analog voltages and allow cascading of multiple layers of MACs.
	To implement a low-energy mixed-signal delay-multiplier, major challenge is the design of a tune-able delay-cell, having linear transfer characteristics. Miyashita et al. \cite{Miyashita2014AnProcessing} first proposed the use of analog-digital mixed signal delay-MAC. Common mathematical operations like addition, subtraction, multiplication, and max-/minimization were demonstrated in a clocked time-domain. However, the multiplication using clocked time-to-digital converters negated power-savings expected from an analog processor. Clock-less tune-able delay-cells for MACs were later proposed in \cite{Miyashita2017AProcessing}, where the accumulation after each dot-product in a binary convolutional neural network was carried implicitly by the delays of a series of nMOS resistor-based delay-cells. However, the use of resistors for enabling scaling of delay, lead to an area-expensive solution. 
	\par
	Delay-modulation via programming voltages, for both $-$ low-power front-end analog processing and approximate-computing acceleration, was demonstrated in \cite{Gopal2018ACMOS}. Applying a small-signal analog input to the back-gate of the transistor modulated the threshold voltage and hence, the delay. Since the threshold voltage varies with the input in a square-root fashion, the delay is inherently non-linear. Also, variation in the threshold voltage across a chip can introduce non-homogeneity in the multiplier. 
	\par
\begin{figure}[t]
		\includegraphics[scale=0.8]{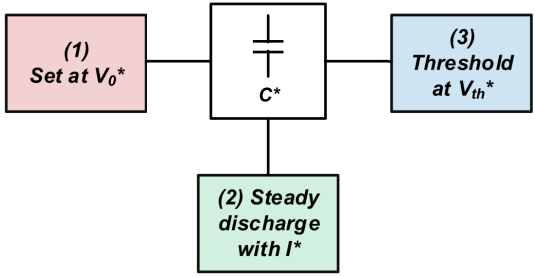}
		\centering
		\caption{Sub-processes of the delay-cell}
		\label{concept1}
	\end{figure}

 
	In this work, a novel CMOS referential delay-cell, based on a steady discharge of a MOSCAP ($C^*$) via a constant current ($I^*$), is proposed. 
	Block-diagram in Fig. \ref{concept1} depicts three sequential processes that $C^*$ undergoes, from $t=0$: 
	\begin{enumerate}
		\item instantaneous pre-charge to $V^*_0$ (colored red) 
		\item steady discharge, through a constant current $I^*$  (blue) 
		\item thresholding of $V^*$ at $V^*_{th}$, using a threshold detector (green).
	\end{enumerate}
	
	With these three processes, time taken for $V^*$ to reach $V^*_{th}$ is:
	\begin{equation}
	T_d=\frac{C^*}{I^*}(V_0^*-V^*_{th}).
	\end{equation}
	If $V_0$ is a linear function of $V_A$, then time taken to discharge to $V^*_{th}$ (or simply, the delay) becomes a linear function of $V_A$. This forms the basis of the proposed delay-cell.
	For use within a multiplier, its delay is exponentially scaled through gate-voltages of the source of $I^*$, rather than transistor widths. 
	Next, analytical models for all the sub-processes in the delay-cell are developed, key sources of jitter identified and a model for the net jitter is formed. From these models, constraints on $C^*$ and $I^*$ for the usability of delay-cells in a multiplier are found, and it's shown that the multiplier cannot accommodate more than 5 bits of (signed) digital-input. Biasing circuits that generate the gate-voltages to scale $I^*$ and the delay exponentially, are then proposed and validated.
	
	\par
	The paper is divided as follows: 
	in Sec. \ref{sec_backg}, necessary but brief background on mixed-signal delay multipliers, delay-MACs and how multiple delay-cells together constitute a multiplier, is presented.
	In Sec. \ref{sec_dc_cmos}, the concept behind the proposed delay cell is presented in more details. For each of the three sub-processes, CMOS implementation details are presented and constraints for a linear delay transfer characteristics developed. Jitter-models are then developed for the two sub-processes that contribute most to the jitter.
	In Sec. \ref{sec_mult}, the constraints and jitter model developed are employed to find the minimum latency and maximum number of bits that can be accommodated within the multiplier. Also, a biasing circuit that enables an accurate exponential scaling of delays of cells within a multiplier is presented.
	In Sec. \ref{sec_discuss}, we discuss about the chosen noise-floor and its connection with the maximum number of bits, and input dependence of energy consumption. Next, the proposed delay-cell is compared with the state-of-art, before concluding in Sec. \ref{sec_concl}.
	
	\section{Background}
	\label{sec_backg}
	
	\begin{figure}[t]
		\centering
		\includegraphics[scale=0.6]{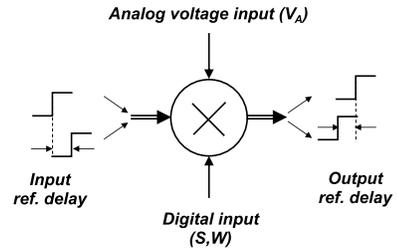}
		\caption{Referential delay}
		\label{backg_0}
	\end{figure}
	
	\begin{figure*}[t]
		\centering
		\subfloat[1-bit]{%
			\includegraphics[scale=0.45]{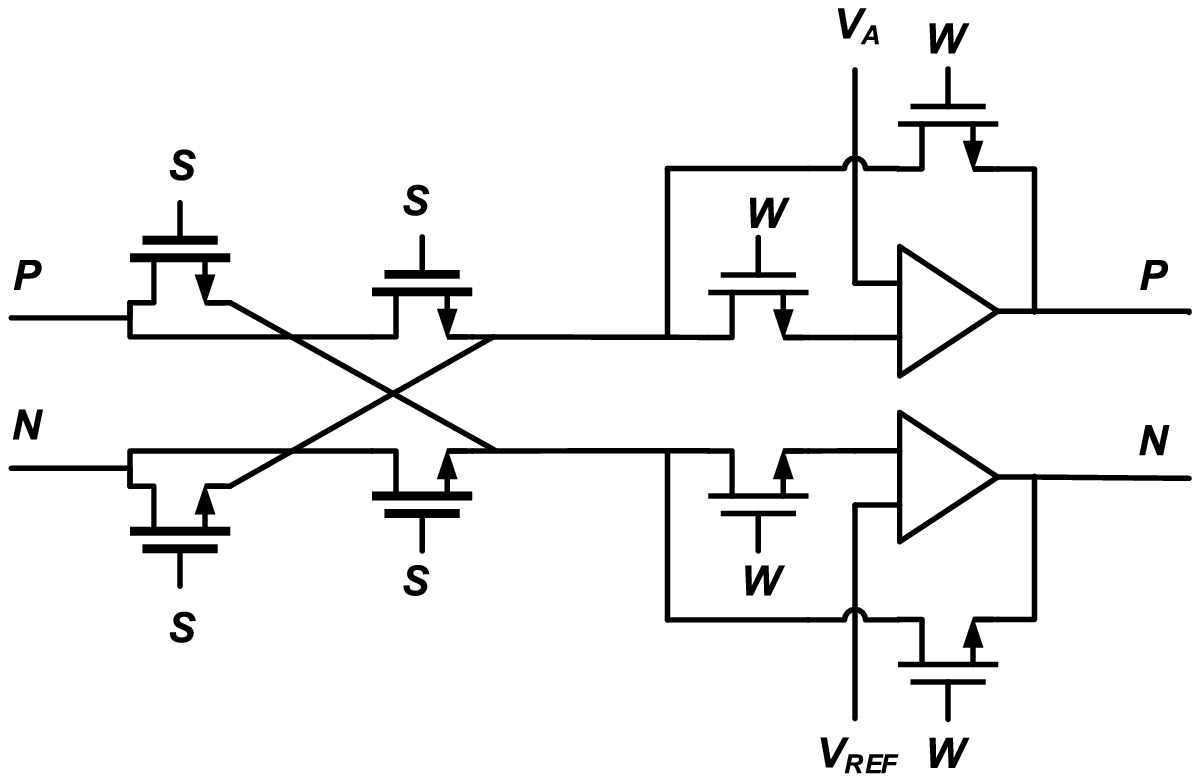}%
		}
		\hspace{0.05\linewidth}
		\subfloat[3-bit]{%
			\includegraphics[scale=0.5]{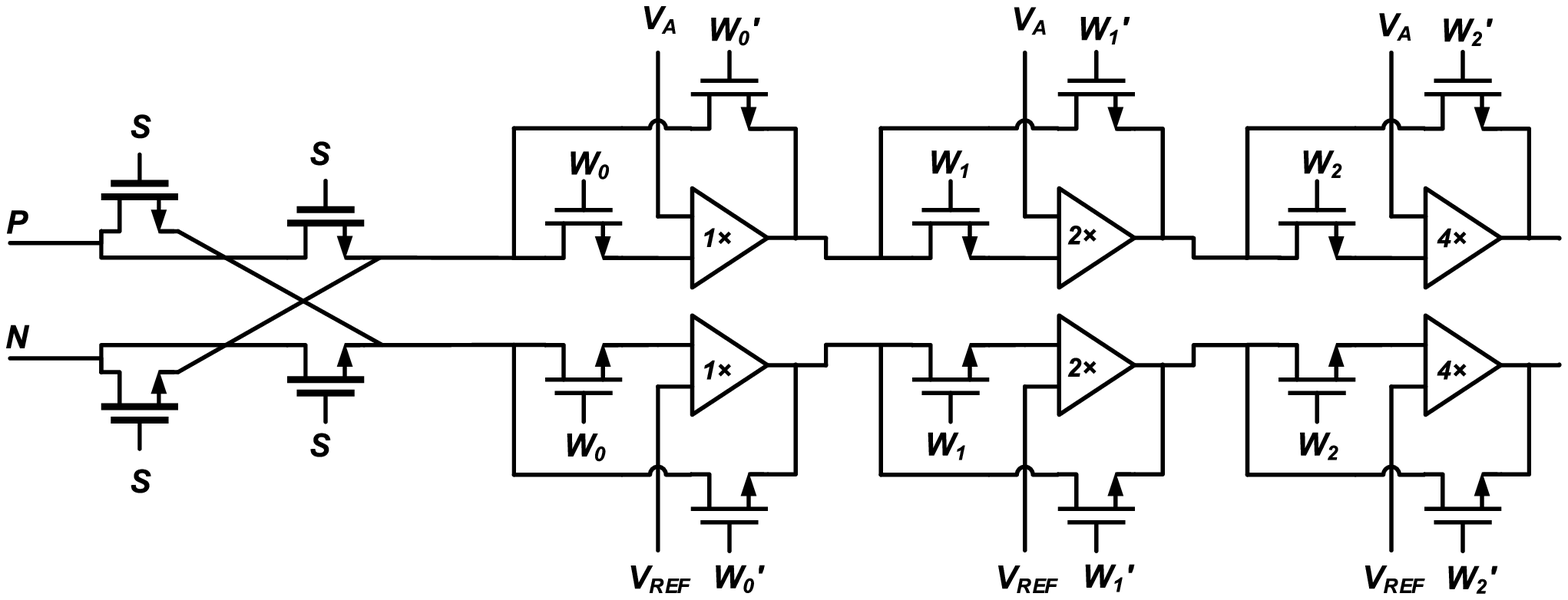}%
		}
		\caption{Signed mixed-signal delay multiplier}
		\label{arch_mult}
	\end{figure*}
	
	A delay-MAC comprises of several delay multipliers, whose delays serially accumulate, before undergoing further non-linear processing.
	The multiplier accepts a time-referenced event signal, which it propagates forward as-is, but after a delay in proportion to the product of its arguments.
	When several such multipliers are placed in series, and a reference event is applied to the first, then, the ref. event is propagated forward, and the net (accumulation of) delay models the dot-product of inputs.
	By using delays, the need of adders is eliminated, because the delays are summed up naturally.
	Since negative numbers cannot be represented using individual events, a pair of events is used, where, the time of instance of one's occurrence referred to the other's, is called referential delay (Fig. \ref{backg_0}).
	\par
	A delay multiplier, besides two input arguments, has a pair of a variable and a reference event-signals (henceforth called \emph{referential} events) at its input and output. For a mixed-signal multiplier, a signed, fixed-point weight vector ($S$ and an n-bit wide vector $W$) and an analog scalar ($V_A$) form the argument, and a pair of rising (or, falling) edges of voltages form referential event-signals (Fig. \ref{backg_0}). To accommodate negative weights, symmetric 2:2 multiplexers (or, relays) are placed within each multiplier that are realized using transmission-gates.  The relay ensures that for each negative weight, the referential events are swapped before multiplication (Fig. \ref{arch_mult}a,b).
	\par Each multiplier consists of smaller referential delay-cells that correspond to each bit of $W$ and create a referential delay equaling $2^i D$, where, $D$ is the common delay-factor and $i\in \{{0,1,2…n-1}\}$. The common delay-factor, $D$, is a linear function of $V_A$. For reasons explained Sec. \ref{subsec_vi}, two, $V_{DD}\rightarrow 0$ falling edges, as referential event signals, are propagated through the multiplier. Weight bits, $w_i$, individually determine whether reference-event signals are delayed by $2^iD$ or not, by making the falling-edge pass or skip a delay-cell. 
	\par 
	Each referential delay-cell has a pair of identical and parallel, linearly tunable delay-cells. One delay-cell inputs $V_A$ and outputs the falling-edge after delay linearly dependent on $V_A$. The second cell inputs a constant reference voltage, $V_{A0}$, and outputs the event after a fixed time. If $V_A>V_{A0}$, then variable event gets more delayed compared to the reference, which represents a positive partial sum. A negative partial sum is produced if $V_A<V_{A0}$ and zero, if $V_A=V_{A0}$. Thus, using a pair of delay-cells homogenises the multiplier with respect to the multiplicand and allows negative weights and referential delays.
	\par
	For illustration, a 1-bit multiplier is shown in Fig. \ref{arch_mult}a. The multiplier comprises of a 2:2 relay and a referential delay cell. $S=1$ implies a negative weight which causes the falling-edges to get swapped. The referential delay-cell further contains two delay-cells, with one variable input ($V_A$) and the other with a reference input ($V_{A0}$). When $W=w_0=0$, cell-bypassing MUX is enabled leading to both negligible delay and ref. delay. Next, 3-bit multiplier is shown in Fig. \ref{arch_mult}b. The referential delays are scaled in the ratio 1,2 and 4, by scaling the absolute delays in the ratio 1,2 and 4. 
	
	One may also use differential mode of operations, where, the referential-delay cell is replaced with differential delay-cell. In this mode, the analog input are changed from $V_A$ and $V_{A0}$ to $V_{A0}+V_{A}$ and $V_{A0}-V_A$. This may remove second order distortion terms of the multiplier without changing the multiplier's circuitry.
	
	\section{Delay-cell Design in CMOS}
	\label{sec_dc_cmos}
	
	\subsection{Steady discharge-based delay-cells}
	
	\begin{figure}[t]
		\includegraphics[scale=0.6]{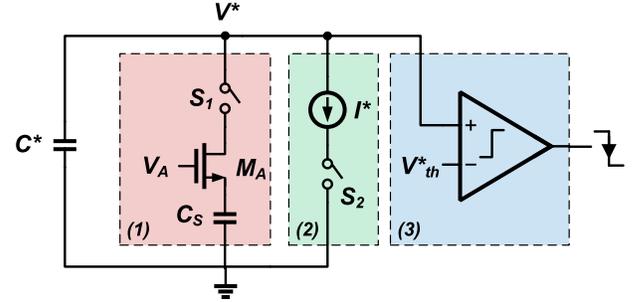}
		\centering
		\caption{Three sub-processes with an n-FET for the initial discharge's linearity}
		\label{concept2}
	\end{figure}
	
	\begin{figure}[t]
		\centering
		\subfloat[Varying $V_A$]{%
			\includegraphics[width=0.5\linewidth]{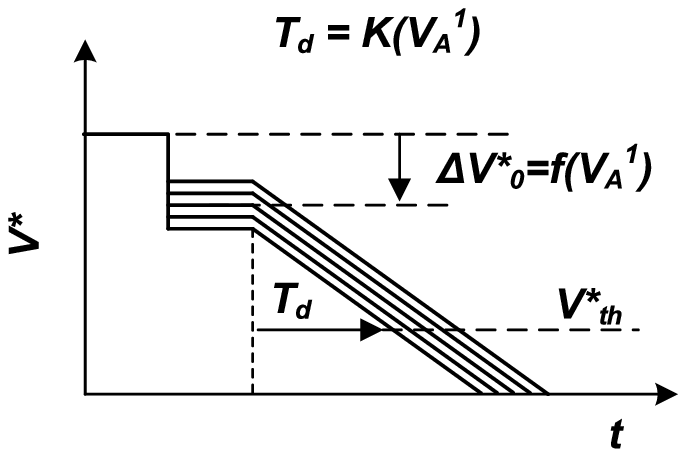}%
		}
		\subfloat[Varying $I^*$]{%
			\includegraphics[width=0.5\linewidth]{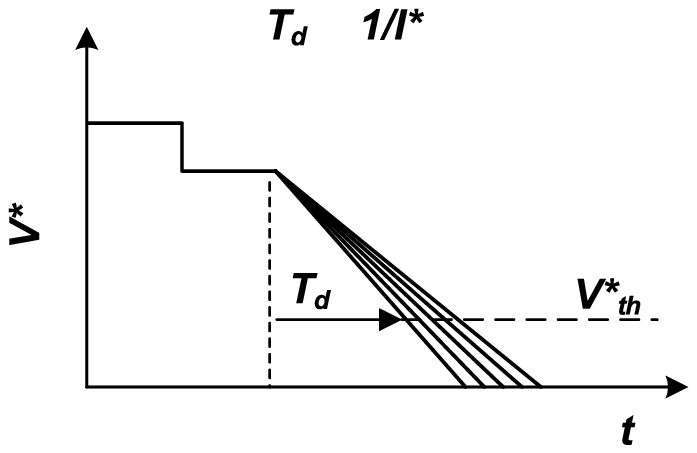}%
		}
		\caption{Expected transient response}
		\label{tranmodel}
	\end{figure}
	
	\begin{figure*}[t]
		\centering
		\includegraphics[scale=0.6]{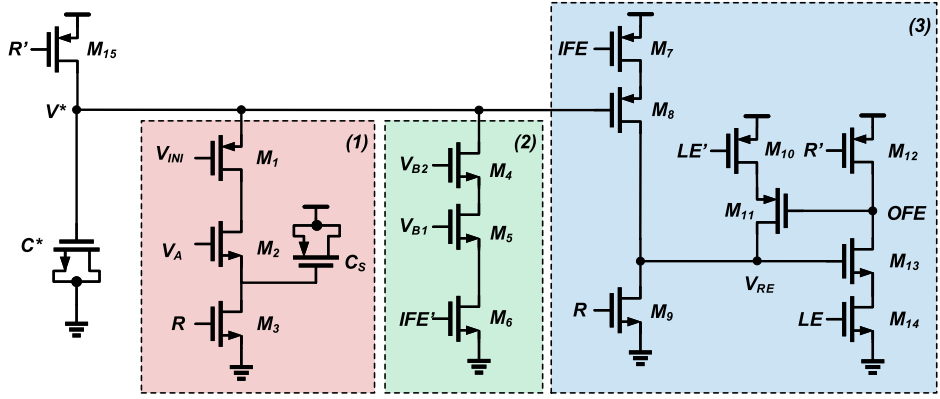}
		\caption{Delay-cell schematic. All transistors have minimum widths}
		\label{dcs}
	\end{figure*}
	
	An idealized circuit implementing this process's equivalent is shown in Fig. \ref{concept2}, where, each component responsible for the three sub-processes have been boxed and colored correspondingly.
	
	In branch 1, the key component is a $V_A$-accepting $M_A$ that has a net source-capacitance $C_S$. Initially, both $S_1$ and $S_2$ are open and $C^*$ is charged to $V_{DD}$. Once $S_1$ is closed, NFET initializes $C^*$ by sinking its charge into $C_S$, until its source-voltage reaches approximately $V_{thn}$ below the gate voltage, to $V_A-V_{thn}$. From charge conservation, the $V^*$ lowers by:
	\begin{equation}
	\label{eq_dv_va_approx}
	{\Delta}V^*_0\approx\frac{C_S}{C^*}(V_A-V_{thn}).
	\end{equation}
	
	Thus, the n-FET conducts until the source voltage rises enough to cut-off the channel, establishing a linear relationship between $V_A$ and ${\Delta}V^*_0$. The approximation in Eq. \ref{eq_dv_va_approx} comes from the fact that a real sub-micron FET doesn't have a well defined threshold-voltage. However, as later shown in Sec. \ref{subsec_vi}, the linear relationship still holds well if $V_A>V_{thn}$.
	\par
	Once the voltage across $C_S$ is set, $S_1$ is opened and $S_2$ is closed causing $C^*$ to spontaneously discharge  via $I^*$, at a constant rate (Fig. \ref{concept2}). The steady discharge process can be described as:
	\begin{equation}
	\Delta V^* (t)={\Delta}V^*_0+\frac{I^*}{C^*} (t-t_{0}),
	\end{equation}
	where, $\Delta V^*$ is the drop in $V^*$ below $V_{DD}$. Next, a threshold-detector, with a threshold $V^*_{th}$, outputs a falling-edge once $V^*$ drops below $V^*_{th}$. Time taken for $\Delta V^*$ ($=V_{DD}-V^*$) to reach a given threshold $\Delta V_{th}^*$  $(=V_{DD}-V^*_{th})$, called the absolute delay ($T_d$), is given by:
	
	\begin{equation}
	\begin{split}
	\label{eq_absdelay}
	T_d &= \frac{C^*}{I^*}\left[\Delta V^*_{th} - \Delta V^*_0\right] \\
	&\approx\frac{C^*}{I^*}\left[\Delta V^*_{th} - \frac{C_S}{C^*}(V_A-V_{thn})\right]
	\end{split}
	\end{equation}
	
	This is a linear function of $V_A$. So, the delay can be adjusted linearly with the input (Fig. \ref{tranmodel}). As discussed in \ref{sec_backg}, to homogenize the delay-input relationship, referential delay is used. For a pair of steady discharge based delay-cells, the referential delay $\Delta t_D$ is:
	
	\begin{equation}
	\label{eq_refdelay}
	\begin{split}
	\Delta t_D&=T_d-T_{d,REF}\\
	&=- \frac{C_S}{I^*}(V_A-V_{A0}),
	\end{split}
	\end{equation}
	which, is independent of $C^*$. In case $C^*$ varies with $V^*$, the referential delay may be written as:
	
	\begin{equation}
	\Delta t_D=\frac{1}{I^*}\int_{V^*_{0}(V_{A0})}^{V^*_0(V_A)}C(v)dv.
	\end{equation}
	
	For $\Delta t_D$ to be a linear function of $V_A$, $C(V^*)$ needs to be maximally constant for the range of $V^*\in\left[V^*_0,V^*_{th}\right]$. If $I^*$ is scaled (exponentially) by factor of $2$ (Fig. \ref{tranmodel}), then a delay multiplier with a digital input vector $\bar{W}=\{w_i, \forall i=1,2,...n\}$ will yield the following referential delay:
	\begin{equation}
	\Delta t_D=\sum_{i=1}^{n}\frac{w_i}{2^n}\left[\Delta t_{D,i}(V_A)\right],
	\end{equation}
	where, $\Delta t_{D,i}$ is the ref. delay from the delay cell \emph{i}. This forms the basis of our mixed-signal delay multiplier.
	\par 
	The schematic of the delay-cell in CMOS is given in Fig. \ref{dcs}, detailed design methodology of which, is discussed next.
	
	\subsection{CMOS implementation: $C^*$}
\begin{figure}[t]
		\centering
		\includegraphics[scale=0.85]{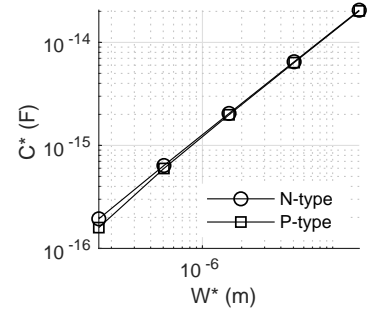}
		\caption{$C^*$ vs. FET width, $W^*$}
		\label{cap}
	\end{figure}

	Later in Sec. \ref{subsec_opt}, it is shown that $C^*$ of approximately $2 fF$ is optimal for minimizing energy, latency and jitter. An inverted MOSCAP, steadily discharging towards depletion, reliably provides capacitance in this range. Since an n-type MOSCAP has a larger inversion capacitance-density than p-type (Fig. \ref{cap}), the prior is used.
	
	\subsection{CMOS implementation: Voltage initialization}
	\label{subsec_vi}
	This stage comprises of min. sized transistors $M_{1-3}$ and pMOSCAP $C_S$ in Fig. \ref{dcs}, key design considerations of which are discussed next:
	\begin{enumerate}
		
		\item \emph{Input nFET $M_A$ ($M_{2}$)}:
		To nullify effect of process variations (PV), the $V_A$-accepting nFET is unique to a multiplier, i.e. it is shared by all delay-cells within a multiplier. Specifically, random dopant-fluctuation, oxide-thickness variations and other process-related non-idealities may offset $V_{thn}$, that may in-turn offset output ref. delay by: 
		\begin{equation}
		\Delta t_D=- \frac{C_S}{I^*}\Delta V_{thn},
		\end{equation}
		
		where, $\Delta V_{thn}$ models effect of PV
		\item \emph{$C_S$}:
		This capacitor is unique to a delay-cell, and a min. sized pFET is assigned to each cell. The pFET stays in the inversion regime regardless of  $V_A$, because $V_S$ saturates to a value that is at least $V_{thn}$ less than $V_{A,max}(=V_{DD})$
		
		\item \emph{Switch S-1 ($M_1$)}:
		In the relevant regime of operation, $V^*$ remains close to $V_{DD}$, necessitating a p-type FET. The switch is unique to each delay-cell, as it isolates $C^*$ of each cell from a shared $M_2$ of the multiplier
		\item \emph{Reset switch ($M_3$)}:
		A switch to reset the $V_S$ to zero before each computation is kept common to all cells within a multiplier
	\end{enumerate}
	
	\par
	If the initial charge on $C_S$ is zero, then, $\Delta V^*_0$ can be expressed as a linear function of $V_A$ and an offset term:
	
	\begin{equation}
	\label{eq_vimodel}
	\begin{split}
	\Delta V^*(V_A)&=\frac{C_{S}+C_{pS}}{C^*}(V_A-V_{thn})+\frac{\Delta Q_{of}}{C^*}\\
	&\approx\frac{C_{S}}{C^*}(V_A-V_{thn})+\frac{\Delta Q_{of}}{C^*}.
	\end{split}
	\end{equation}
	
	Here, $C_{pS}$ is the parasitic capacitors, arising from $M_2$ and $M_3$; $\Delta Q_{of}$ models the zero-offset at $V_A= V_{thn}$ dependent on several parameters: $C^*$, $M_2$'s width and other parasitic effects like feed-forward of input falling-edge into $C^*$. Note that $\Delta Q_{of}$ has two distinct values: first is defined within the discharge-pulse application (MD) and it contains a feed-forward component of the falling-edge. The second is defined after the discharge-pulse application (PD), and is slightly less than MD.
	\par
\begin{figure}[t]
		\centering
		\includegraphics[scale=0.75]{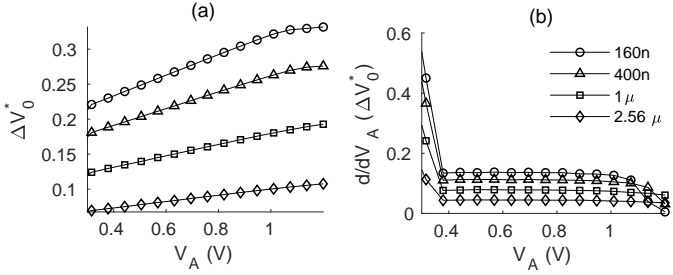}
		\caption{$\Delta V^*$ plots, vs. $V_A$ (a) Absolute value (b) Derivative}
		\label{init_1}
	\end{figure}

	To quantify linearity, $\Delta V^*$ is plotted in Fig. \ref{init_1} against $V_A$ and its derivative w.r.t. $V_A$ in \ref{init_1}b for $V_A$ ranging between $0.3V$ and $1.2V$, with $W^*$ (or $C^*$) as parameters of design.  For this range, less than 10\% variation is seen. The figure shows that larger capacitors can provide better linearity.
	\par
\begin{figure}[t]
		\centering
		\subfloat[Deriv. $\Delta V^*$ vs. $W^*$]{%
			\includegraphics[scale=0.68]{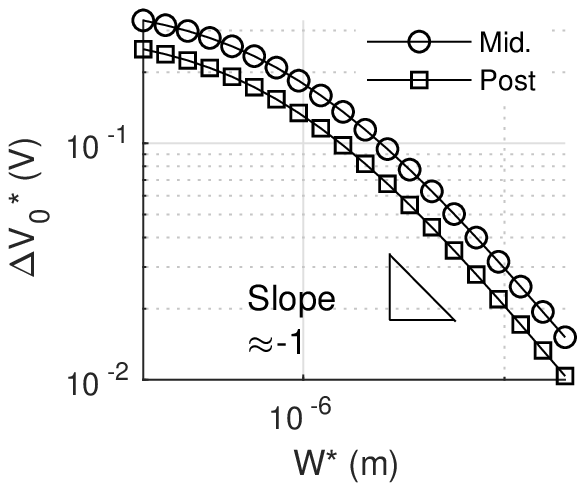}%
		}
		\subfloat[$\Delta V^*$ vs. $W^*$]{%
			\includegraphics[scale=0.68]{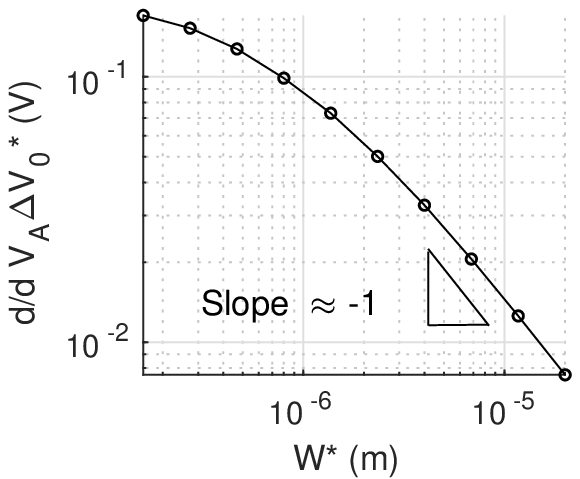}%
		}
		\caption{Variation of $\Delta V^*$ with $W^*$ and linearization models}
		\label{init_2}
	\end{figure}

	Fig. \ref{init_2} plots the  $\Delta V^*_{0}$ for and its average derivative, versus $W^*$ ($\propto C^*$), in a \emph{log-log} fashion. Both plots have a constant slope of $-1$ for sufficiently large $W^*$, validating Eq. \ref{eq_vimodel} as a model for the discharge process. $C_S+C_{pS}$ and $\Delta Q_{of}$ are then empirically determined by fitting the model of Eq. \ref{eq_vimodel}, yielding $C_S+C_{pS}=0.23 fF$ and $\Delta Q_{of} =0.5 fC$ (MD).
	
	Eq. \ref{eq_vimodel} is only valid when the $C^*$'s voltage is big enough to charge up $C_S$. Mathematically,
	\begin{equation}
	V_{DD}-\Delta V^*(V_A)>V_A-V_{thn}
	\end{equation}
	
	For $V_A=V_{DD}$, we get:
	\begin{equation}
	\label{eq_constraint_vi}
	\Rightarrow C^*>\frac{(C_S+C_{pS})(V_{DD}-V_{thn})+\Delta Q_{of}}{V_{thn}}
	\end{equation}
	
	This sets the lower limit on $C^*$, which is employed later in Sec. \ref{subsec_opt}.
	\par
	To slightly enhance the linearity without adding to the area, one in every 5 pMOSCAP of the $C_S$ is replaced by an nMOSCAP. For $V_S<V_{DD}-V_{thp}$, PFET is inverted and provides a close to a  constant cap. For $V_S>V_{DD}-V_{thp}$, pFET's capacitance diminishes but nFET offsets the loss. Since NMOS is smaller,  it does so, only to a small extent. 
	
	\subsection{CMOS implementation: Steady discharge}
	\begin{figure}[t]
		\centering
		\subfloat[$V^*$ transient (simulated)]{%
			\includegraphics[scale=0.6]{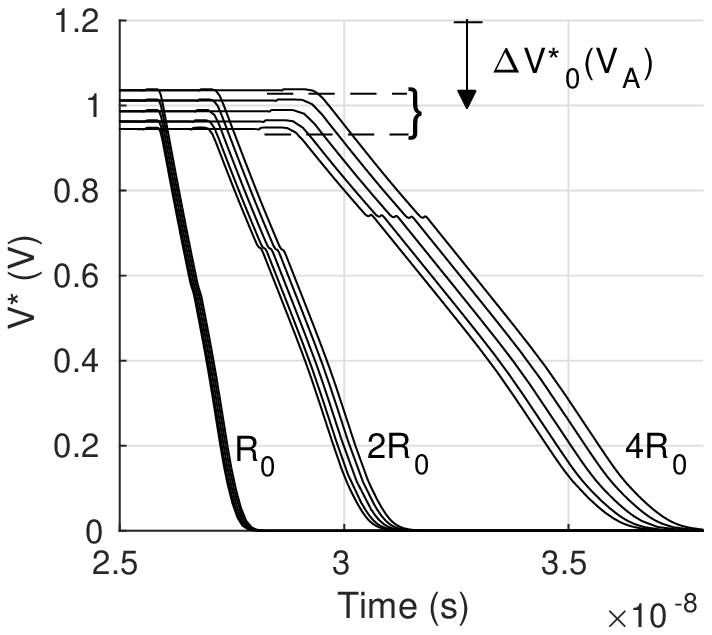}
		}
		\subfloat[ $T_d$ at $V_{DD}/2$ vs. $V_A$ ]{%
			\includegraphics[scale=0.55]{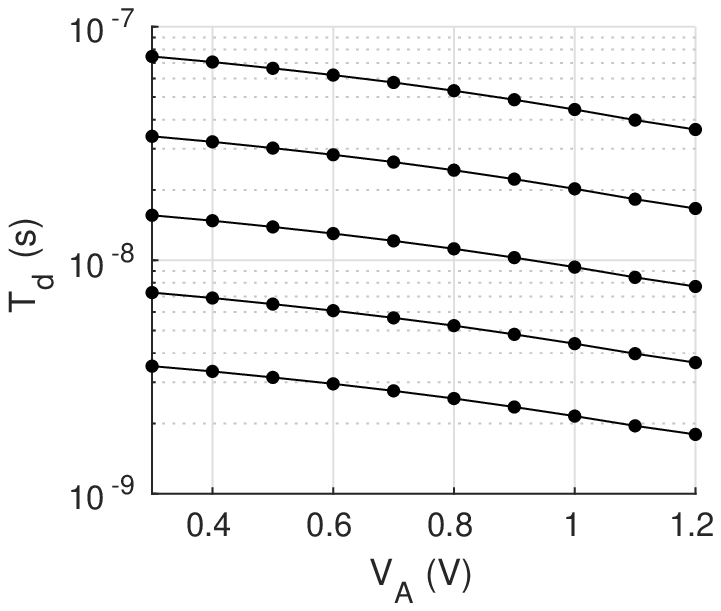}
		}
		\caption{Steady discharge's characterization}
		\label{steady_disch}
	\end{figure}
	
	\label{subsec_sd}
	This part of the delay-cell consists of transistors $M_{4-6}$ in Fig. \ref{dcs}, key design considerations of which are discussed next:
	\begin{enumerate}
		\item \emph{$I^* (M_4 - M_5)$}:
		This is realized using bi-cascoded nFET current source, with the FETs at their min. widths. The exponential current scaling is done via an external biasing circuit, that generates gate-voltages for both $M_4$ and $M_5$. The biasing circuits are discussed in Sec. \ref{subsec_biasckt}.
		
		\item \emph{Switch S-2 ($M_6$)}:
		An NMOS switch is placed in series with the current-source. Unlike S-1, the switch was placed away from $C^*$, preventing the feed-forward through the parasitic capacitors.
	\end{enumerate}
	
	For $W_S$ ($\propto C_S$) at its minimum value ($160 nm$) and $W^* (\propto C^*)=640 nm$, $C^*$’s discharge transient is shown in Fig. \ref{steady_disch}a. Referring back to Fig. \ref{dcs}, the switch ($M_2$) is turned ON at $t=25 ns$, by ramping-up $IFE'$, the input to $M_6$. As expected, $I^*$ discharges $C^*$ at a near-constant rate (Fig. \ref{steady_disch}a). Its constancy depends solely on output resistance of the current source.
	Kinks observable in the transients are caused by the feedback from the half-latch and do not practically affect the performance.
	The absolute delay for various $R$ ($=I^*/C^*$), spaced exponentially with a factor of 2, is plotted in Fig. \ref{steady_disch}b. 
	
	\subsection{CMOS Implementation: Threshold detector}
	\label{subsec_TD}
	It comprises of $M_{7-14}$ as the falling-edge, uni-polar threshold detectors, half-latch and other switches for resetting. Details and design consideration are discussed next:
	
	\begin{enumerate}
		\item \emph{Falling-edge inverter ($M_8$)}: To minimize the area requirements, the width of $M_8$ is kept minimum. As discussed below, under certain constraints on $C^*$, this inverter contributes to a $V_A$-independent delay, thus keeping distortion negligible
		\item \emph{Leak-prevention switch ($M_{7}$)}: It prevents the sub-threshold $M_8$ from leaking and set-up the latch pre-maturely. It inputs the falling-edge of the previous delay cell
		\item \emph{Half-latching inverter ($M_{11,13}$)}: These transistors latch $V_{RE}$ to $V_{DD}$ and the $OFE$-node to $0$, once $V_{RE}$ reaches $V_{thn}$
		\item \emph{Latch-en-/disable switches ($M_{10}, M_{14}$)}: These switches enable the half-latch operation when closed and otherwise, disable it, reducing the energy required to reset the half-latch
		\item \emph{Reset-FETs ($M_{9,12}$)}: $M_{10}$ resets node $V_{RE}$ to 0. $M_{12}$ sets the OFE node to $V_{DD}$ before the start of computation. These transistors are shared within the multiplier
		
	\end{enumerate}
	\par
	A CMOS inverter can serve as a low-energy threshold detector, whose switching-voltage can be set by designing the ratio of sizes of pMOS and nMOS. However, it consumes short-circuit energy ($E_{SC}$) given by: 
	\begin{equation}
	E_{SC}=\frac{V_{DD}}{6R} \mu\frac{W}{L} C_{ox}(V_{DD}-V_{thp}-V_{thn})^3
	\end{equation}
	
	where $R=I^*/C^*$ and,
	\begin{equation*}
	\mu\frac{W}{L}=\frac{\mu_p\mu_n(W/L)_p(W/L)_n}{((\mu_p(W/L)_p)^{1/2}+(\mu_n(W/L)_n)^{1/2})^2}.
	\end{equation*}
	
	Since $R$ decreases exponentially, $E_{SC}$ increases exponentially. Hence, a delay-cell implementing the n-th exponent will expend $2^n\times$ the $E_{SC}$ of the cell implementing the first. For a n-bit  multiplier, the total short-circuit energy lost is $\left(2^{n+1}-1\right)E_{SC}$. Thus, an exponential requirement in energy consumption motivates an alternative inverting mechanism.
	\par 
	The low-energy alternative to the CMOS inverter is a standalone pFET ($M_8$ in Fig. \ref{dcs}), due to its switch-like I-V relationship. If it were an ideal switch, with a switching voltage $V_S$ $ (>{\Delta}V^*_{0,max}$), $V_{RE}$ would jump to $V_{DD}$ after a fixed delay following $\Delta V^*(t)=V_{thp}$. This would conserve the linearity of Eq. \ref{eq_refdelay} with respect to $V_A$, as it only adds a constant delay. However, a real PFET has a close to exponential I-V relationship and conservation of linearity needs to be established, or at least constraints for maximal linearity determined. 
	\par
	With the assumption of exponential I-V characteristics and large output-resistance ($g_{DS}$), the sub-threshold current can be expressed as a function of the gate-source voltage ($=\Delta V^*_0$) using the following equation:
	\begin{equation}
	\label{eq_fet_iv}
	I=I_0\exp\left(\frac{\Delta V^*_0}{V_T}\right),
	\end{equation}
	where, $V_T$ is the thermal voltage. Eq. \ref{eq_fet_iv} is valid only for $\Delta V^*_0<V_{thp}$; for $V_{GS}>V_{thp}$, I-V relationship is usually degree-2 or less polynomial, moving the switch away from an ideal behavior. 
	\par
	For a $V_A$ that linearly decreases from ${\Delta}V^*_0$ with a steady rate $R$, $V_{RE}$ (Fig. \ref{dcs}) can be expresses as a function of time using:
	\begin{equation}
	\label{eq_vre_tranmodel}
	V_{RE}(t)=\frac{I_0}{C}\frac{V_T}{R}\exp\left(\frac{{\Delta}V^*_0}{V_T}\right)\left(\exp\left(\frac{Rt}{V_T}\right)-1\right),
	\end{equation}
	where, $R=\frac{I^*}{C^*}$ is the rate of change of $V^*$ with time, and $C$ is net capacitance at the drain of M8. When $V_{RE}(t)=V_{thn}$, the half-latch is set up and the voltage at $OFE$ node (Fig. \ref{dcs}) falls to $0$. Thus, the time taken from the start of discharge ($t=0$) to the drop in $OFE$-node voltage to zero $\left(t=T_d\right)$ is:
	\begin{equation}
	T_d=\frac{V_T}{R}\ln\left(R\frac{CV_{thn}}{I_0 V_T}\exp\left(-\frac{{\Delta}V^*_0}{V_T}\right)+1\right).
	\end{equation}
	$T_d$ becomes a linear function of ${\Delta}V^*_0$ under the constraint:
	\begin{equation}
	\frac{RCV_{thn}}{I_0 V_T}\exp\left(-\frac{{\Delta}V^*_0}{V_T}\right) \gg 1
	\end{equation}
	Putting $R=\frac{I^*}{C^*}$, this inequality may alternatively be written as:
	\begin{equation}
	\label{eq_td_constraint}
	\frac{I^*}{C^*}\frac{C}{I_0\exp\left(\frac{{\Delta}V^*_0}{V_T}\right) }\frac{V_{thn}}{V_T}\gg 1
	\end{equation}
	Since ${\Delta}V^*_0$ varies inversely with $C^*$ (from Eq. \ref{eq_vimodel}), the denominator in Eq. \ref{eq_td_constraint} is a monotonically decreasing function of $C^*$. Then, as per this inequality, $C^*$ should be greater than a critical capacitance $C^*_{min}$. This inequality is used in Sec. \ref{subsec_opt}, for establishing constraints on $C^*$ and $n$.
	\par
	Under the validity of this inequality, $T_d$ can be expressed as:
	
	\begin{equation}
	T_d=\frac{V_T}{R}\ln\left(\frac{RCV_{thn}}{I_0 V_T}\right)-\frac{{\Delta}V^*_0}{R},
	\end{equation}
	
	matching the expectation of $T_d$'s linearity with ${\Delta}V^*_0$, or $V_A$. Note that the latch-point, or, the value of $V^*$ when $V_{RE}=V_{thn}$ is a constant, independent of $V^*_0$ (or $V_A$), and expressible as:
	\begin{equation}
	{\Delta}V^*_{th}={\Delta}V^*_0+RT_d=V_T\ln\left(\frac{RCV_{thn}}{I_0V_T}\right).
	\end{equation}
	\par
	Though an exponential I-V characteristics is assumed for $M_8$, in reality, it is exponential only for sub-threshold gate voltages. For devices with power I-V relations, Eq. \ref{eq_vre_tranmodel} is re-derived with the modified I-V, and constraints of Eq. \ref{eq_constraint_vi} re-determined. For $p-1$ power current-voltage relationship,
	\begin{equation}
	\frac{I^*}{C^*}\frac{C}{I(V_A,C^*)}\frac{V_{RE}}{V_{G0}/p}\gg 1.
	\end{equation}
	For an ideal switch ($p\rightarrow \infty$), the constraint is trivially satisfied and TD stage doesn't contribute to distortion. As the I-V relationship of the pFET moves away from step-like behaviour towards linearity ($p\rightarrow0$), ensuring linearity from delay-$V_A$ characteristics becomes harder. 
	
	\subsection{Noise-modelling}
	\label{subsec_noise}
	The delay-cell essentially consists of two current-integrators that accumulate the accompanying noise-current, starting from the arrival of falling-edge ($t=0$) to the latch-up ($t=T_d$). This leads to a net temporal shift in the falling-edge, or a jitter in the output falling-edge.  To enable design of the delay-cell and multiplier, the two jitter components are modeled as a function $C^*$ and $I^*$ (the design variables) and an upper limit on jitter is set, yielding constraints on the design variables and $n$.
	For simplifying jitter-modeling, it is assumed that:
	\begin{enumerate}
		\item Out of the three, only two processes contribute to the jitter: steady discharge and threshold-detection. (Initial discharge occurs much faster than $T_d$, so it contributes negligibly to the net jitter.)
		\item The net jitter is much smaller than $T_d$
	\end{enumerate}
	
	\subsubsection{Jitter from steady discharge}
\begin{figure}[t]
		\centering
		\subfloat[Iso-$I^*$]{%
			\includegraphics[scale=0.7]{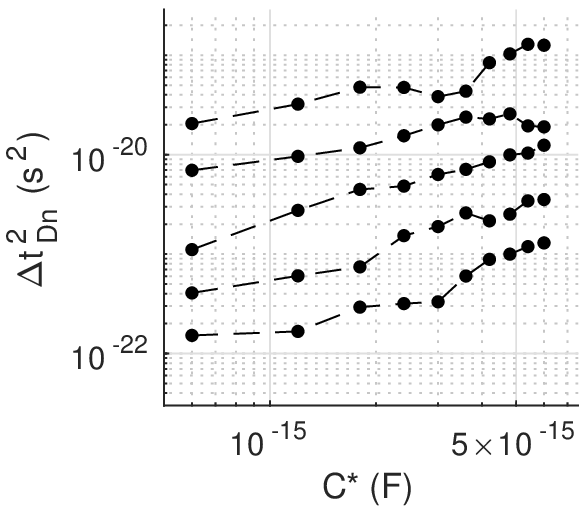}%
		}
		\subfloat[Iso-$C^*$]{%
			\includegraphics[scale=0.7]{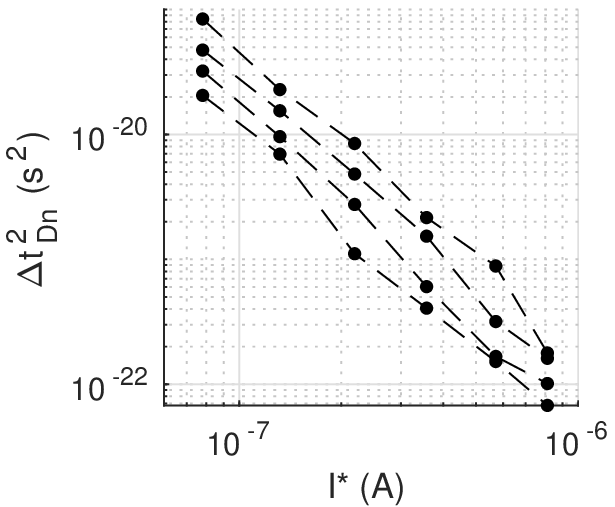}%
		}
		\caption{steady discharge-stage's jitter}
		\label{noise1_plots}
	\end{figure}

	For this stage, the primary contributor of jitter the is channel noise-current from $M_{4,5}$ accumulating in $C^*$.
	To simplify the model, it is assumed that the noise-current out of $M_4$ circulates within itself, and hence contributes negligibly to the jitter. With this assumption, the stage reduces to a noisy FET discharging a fixed capacitor, jitter modelling for which was done for ring oscillators in \cite{Abidi2006PhaseOscillators}.  For an inverter-type ring-oscillator, the jitter-per-stage is modeled as:
	\begin{equation}
	\Delta t_{Dn}^2=\frac{4kT\gamma g_{d0}}{2I^{*2}}T_{d},
	\end{equation}
	where, $\gamma$ is the excess noise factor, $g_{d0}$ is the drain-source conductance at $V_{DS}=0$. This naturally extends to the proposed delay-cell, with the exception that $T_{d}$ is \textit{variable}, dependent on the rate of discharge and $V_A$.
	Using Eq. \ref{eq_absdelay} the expression for jitter becomes:
	\begin{equation}
	\Delta t_{Dn}^2=\frac{4kT\gamma g_{d0}}{2I^{*3}}C^*\left({\Delta}V^*_{th}-{\Delta}V^*_{0}\right)
	\end{equation}
	To further simply, the dependence of jitter on ${\Delta}V^*_{th}$ and ${\Delta}V^*_{0}$ is neglected and a constant jitter, for a $V_{DD}/2$ drop in $V^*$, is defined and used. Owing to the fact that $g_{d0}$ has a linear dependence on current, jitter from this stage is compactly express-able as:
	\begin{equation}
	\label{eq_noise_sd}
	\Delta t_{dn}^2=K\frac{C^*}{I^{*2}},
	\end{equation}
	where, $K$ is a temperature and technology dependent constant. For model validation, the jitter is simulated in software, for IBM's $130nm$ technology. Resulting $\Delta t_{dn}^2$, with only $M_{4-5}$ noise turned on, versus $C^*$ and $I^*$ is plotted in Fig.\ref{noise1_plots}.
	
	Instead of Eq. \ref{eq_noise_sd}, the following model is used as it fits the experimental data better (R-sq. of 0.982, from 10 iterations):
	\begin{equation}
	\Delta t_{dn}^2=K\frac{C^*}{I^{*p}},
	\end{equation}
	
	where, $K=2.95\times 10^{-16}$ and $p=2.46$.
	\\
	
	\subsubsection{Jitter from threshold detector}
	\begin{figure}[!t]
		\centering
		\subfloat[Variance in $V_{RE}$]{%
			\includegraphics[scale=0.7]{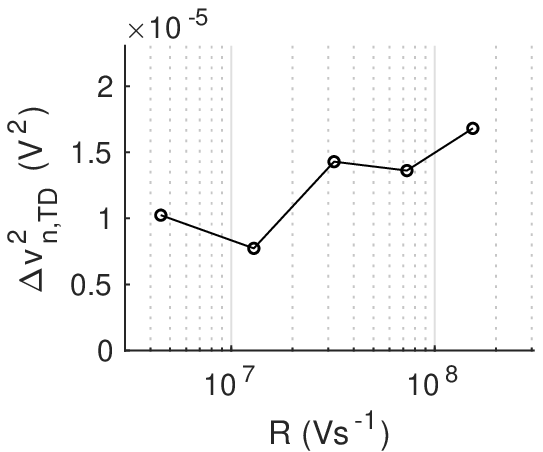}%
		}
		\subfloat[Jitter]{%
			\includegraphics[scale=0.75]{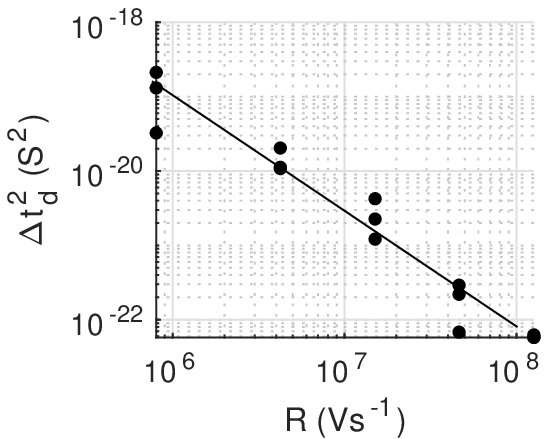}%
		}
		\caption{Variance in $V_{RE}$ and jitter due to $M_8$'s noise-current}
		\label{noise2_plots}
	\end{figure}
	\begin{figure*}[t]%
		\centering
		\includegraphics[width=\textwidth]{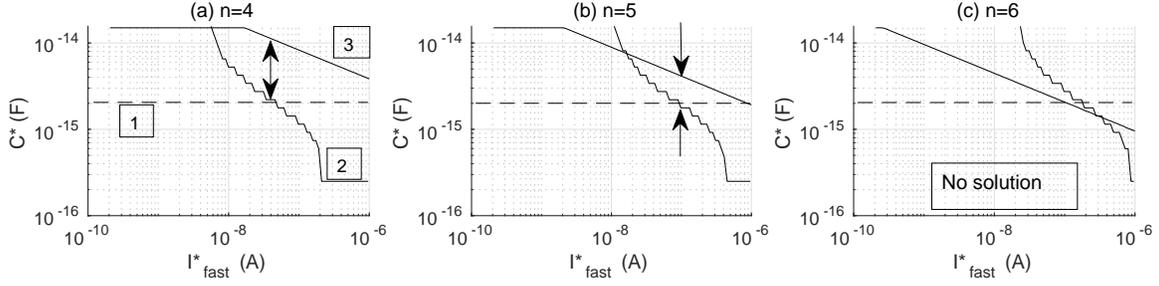}%
		\caption{Constraints 1,2 and 3 for 4,5 and 6 bit multipliers}
		\label{constraints}
	\end{figure*}%
	Since the input gate-source voltage ($\Delta V^*$) of $M_8$ increases linearly with time and drain-current exponentially, it is assumed that RMS channel noise-current $\left(i_n\right)$ out of $M_8$ increases exponentially. Thus, at any given instant of time post falling-edge's arrival, noise current from only the past 3-4 $V_T$-drops in $\Delta V^*$, contributes to this stage's jitter.
	
	If $\Delta v_n$ is the deviation in $V_{RE}$ at $t\rightarrow T_d^-$, then for a constant $i_n$, we have:
	\begin{equation}
	{\Delta}v_n^2\propto \frac{i_n^2}{\Delta f}\frac{T_{dn}}{C^2}. 
	\end{equation}
	Since $i_n^2$ varies exponentially over the duration $t_{dn}$, this equation cannot be applied without adjustments.
	Thus, the following equation is used:
	\begin{equation}
	{\Delta}v_n^2\propto \frac{\int_{0}^{T_d}4kT\gamma g_{d0}(t)dt}{C^2}
	\end{equation}
	Letting $g_{d0}=G_0 \exp\left(\frac{{\Delta}V^*_{0}+Rt}{V_T}\right)$, we get:
	\begin{equation}
	\begin{split}
	{\Delta}v_n^2&\propto \int_{0}^{T_d}4kT\gamma G_0 \exp\left(\frac{{\Delta}V^*_{0}+Rt}{V_T}\right)dt \\
	& = \beta \frac{V_T}{R} \exp\left(\frac{{\Delta}V^*_{th}}{V_T}\right) \\
	& = \beta\frac{CV_{thn}}{I_0},
	\end{split}
	\end{equation}
	where, $\beta=4kT\gamma G_0$. This equation establishes an independence of $v_n$ on $R$, which is confirmed from Fig. \ref{noise2_plots}. In the figure, $R$ is varied by a factor of more than $10\times$, but less than $2\times$ rise is seen in ${\Delta}v_n^2$.  
	\par
	Next, the relationship between $\Delta t_{Dn}$ and $R$ is determined. Similar to the approach adopted in \cite{Abidi2006PhaseOscillators}, $\Delta t_{Dn}$ can be found by extrapolating noisy $V_{RE}$ along the noise-less $V_{RE}$, to the point of latch-up:
	\begin{equation}
	\begin{split}
	\Delta t_{Dn}^2&=\left(\frac{dV_{RE}}{dt}\right)^{-2}{\Delta}v_{n}^2, \\
	&\propto\frac{v_{n}^2}{R^2},
	\end{split}
	\end{equation}
	where, Eq. \ref{eq_vre_tranmodel} was used for ${dV_{RE}}/{dt}\propto R$. Simulated jitter, with only $M_8$'s noise turned on, is plotted in Fig. \ref{noise2_plots}. 
	\par
	For minimally sized $M_{7-14}$, the fitted model from 10 iterations of (noisy) simulations is:
	\begin{equation}
	\Delta t_{dn}^2=K_2 \frac{1}{R^{1.5}},    
	\end{equation}
	where, $K_2=1.29\times 10^{-10}$. Thus, the actual exponent is less than predicted. 

	\section{Mixed-signal Delay Multiplier}
	\label{sec_mult}
	With the delay-cell design considerations discussed, next, the necessary steps to employ the cells within a multiplier are presented: (1) use of constraints to find the valid region of design and operation (2) bias-circuit design for $I^*$'s exponentiation. Lastly, through simulations, the functionality of cascaded delay-cells as multiplier is validated and the key energy components for each multiplication operation are identified.
	
	\subsection{Optimizing $C^*$ and No. of Bits}
	\label{subsec_opt}
	Using the inequalities involving $C^*$, developed in Sec. \ref{subsec_vi}, Sec. \ref{subsec_TD} and the noise models of Sec. \ref{subsec_noise}, the constraints on $C^*$ and $I^*$ are determined. Note that these are valid only for the IBM's $130nm$ technology, but may similarly be determined for other CMOS technology nodes.
	
	\subsubsection{Linearity of voltage-initialization}
	In the inequality of Eq. \ref{eq_constraint_vi}, replacing model-parameters extracted from the data of Fig. \ref{init_1}-\ref{init_2} gives constraint 1,
	\begin{equation}
	\begin{split}
	C^* &> 2.2f,\\
	\end{split}
	\end{equation}
	
	which, corresponds to an inverted nMOSCAP single-finger width of $1.28\mu m$. This constraint is marked by '1' in Fig. \ref{constraints}a-c.
	\subsubsection{Linearity of threshold detection}
	Since $I^*$ of the slowest cell is $2^n$ times smaller than than that of the fastest cell ($I^*_f$), constraint 2 from Eq. \ref{eq_td_constraint} becomes:
	\begin{equation}
	\label{eq_constraint2}
	\frac{2^{-n}I^*_f}{C^*}\frac{C}{I_0\exp\left(\frac{{\Delta}V^*_0(V_A,C^*)}{V_T}\right) }\frac{V_{thn}}{V_T}> 1
	\end{equation}
	
	Only $C^*$ and $I^*_f$ are designable; the rest $-$ $C, V_{thn}, V_T$ and $I_0$, are constant. To simplify the analysis, the denominator is maximized over $V_A$ and the uni-variate $\Delta V^*_0\left(V_A=1.2,C^*\right)$ used.  Though the argument of the exponential in Eq. \ref{eq_constraint2}, $\Delta V_0\left(V_A,C^*\right)$, was modeled in Sec. \ref{subsec_vi}, actual data of Fig. \ref{init_1} is used. This constraint is marked by '2' in Fig. \ref{constraints}a-c.
	
	\subsubsection{Upper limit on jitter}
	\label{subsubsec_jitter_lim}
	The referential delay of the fastest cell, from Eq. \ref{eq_refdelay}, is:
	\begin{equation*}
	\Delta t_D=- \frac{C_S}{I^*_{f}}(V_A-V_{A0})
	\end{equation*}
	
	If jitter from steady discharge is denoted by $\Delta t_{Dn,1}$ and from TD by $\Delta t_{Dn,2}$, the constraint on the net jitter is such that it is to be smaller than the maximum ref. delay of the fastest cell. For a $V_{A0}=0.75V$, 
	\begin{subequations}
		\label{eq_noise_constraint}
		\begin{align}
		3\sqrt{\Delta t_{Dn,1}^2+\Delta t_{Dn,2}^2}\leq 0.4\frac{C_S}{I^*_f} \\
		3\sqrt{K\frac{C^*}{({2^{-n}I^*_f})^{2.46}}+K_2 \left(\frac{C^*}{2^{-n}I^*_f}\right)^{1.5}} \leq 0.4\frac{C_S}{I^*_f}.
		\end{align}
	\end{subequations}
	
	With its LHS being monotonic function of $C^*$, Eq. \ref{eq_noise_constraint} gives an upper limit on $C^*$ for a given $n$. This constraint is marked by '3' in Fig. \ref{constraints}a-c.
	
	\par
	Fig. \ref{constraints} plots the constraints for a 4, 5 and 6-bit multiplier. For 4 and 5 bits, the valid region of operation, marked by double-sided arrows, lies between the curves corresponding to constraints 1, 2, and 3. For 6 bits, no solution exists for the chosen constraints. Thus, the 5-bit multiplier with 
	\begin{equation*}
	C^*=2.2fF,
	\end{equation*}
	
	and 
	\begin{equation*}
	I^*_f=1\mu A
	\end{equation*}
	
	emerges as the point of design, as it works for all multipliers with less than 6 bits, minimizes the multiplication latency and energy consumption.
	
	\subsection{Biasing circuit}
	\label{subsec_biasckt}
	Accurate biasing for the current sources is required to ensure low output distortion. As discussed below, its behavior must meet two specifications: 
	\begin{enumerate}
		\item As discussed in Sec. \ref{subsec_opt}, $I^*_f$ is achievable only for the sub-threshold transistors with the employed VLSI node. Hence, the biasing circuit is designed only for sub-threshold currents and works well in this regime only
		\item The current source within each cell consists of a pair of series NFETs ($M_4$ and $M_4$ in Fig. \ref{dcs}). $M_5$'s gate-voltage (primary bias) is such that it sinks $2^{-n}I^*$ and its drain-voltage is fixed close to $100 mV$ ($\approx4V_T$). The drain voltage is maintained by $M_4$ gated with a secondary bias approximately $100 mV$ above $M_5$ 
	\end{enumerate}

	\par
\begin{figure}[t]
		\centering
		\includegraphics[width=0.7\linewidth]{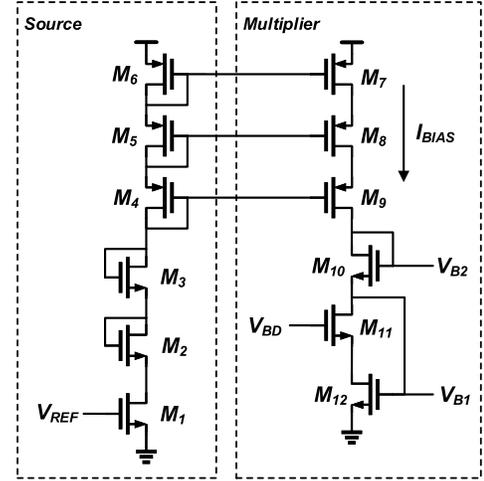}
		\caption{Biasing circuit. $M_{1-6}$ constitute source branch and $M_{7-12}$ constitute the scaling branch}
		\label{bias_schem}
	\end{figure}

	The circuit (Fig. \ref{bias_schem}) has two branches: \emph{source} and \emph{scaling}. Source branch ($M_{1-6}$) generates biasing voltages dependent on a programmable voltage, $V_{REF}$. Scaling branch ($M_{7-12}$) first uses those biasing voltages to generate current in the exponents of 2 (using transistor widths) and then generates the bias for the cells' current-source using self-biasing. Here, the primary bias out of $M_{12}$ is denoted as $V_{B1}$ and the secondary out of $M_{10}$ as $V_{B2}$.
	
	\par
	In the source branch, $M_1$ converts the reference voltage $V_{REF}$ into current $I_{BIAS}=I^*_{f}$. $M_2$ and $M_3$, being self-biased in the saturation regime, push up the gate voltages of the tri-cascode, enough to keep mirror transistors ($M_{10-11}$) of the scaling branch saturated. $M_{4-6}$ produce the multiplier-cascode's  bias. 
	
	\par
\begin{figure}[t]
		\centering
		\subfloat[Source branch's current]{%
			\includegraphics[width=0.45\linewidth]{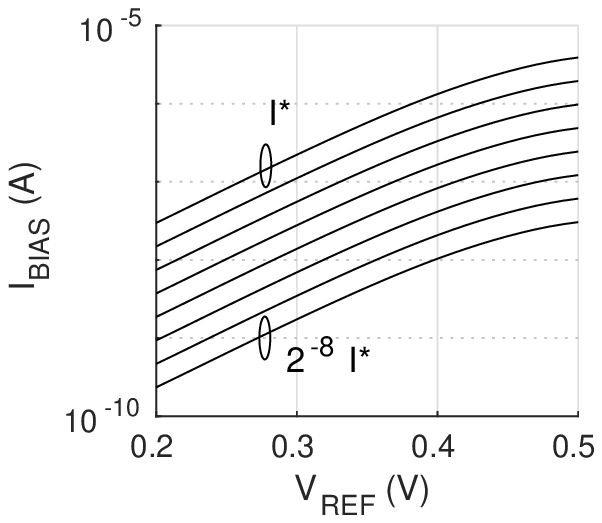}%
		}
		\subfloat[Output $V_{B1}$]{%
			\includegraphics[width=0.45\linewidth]{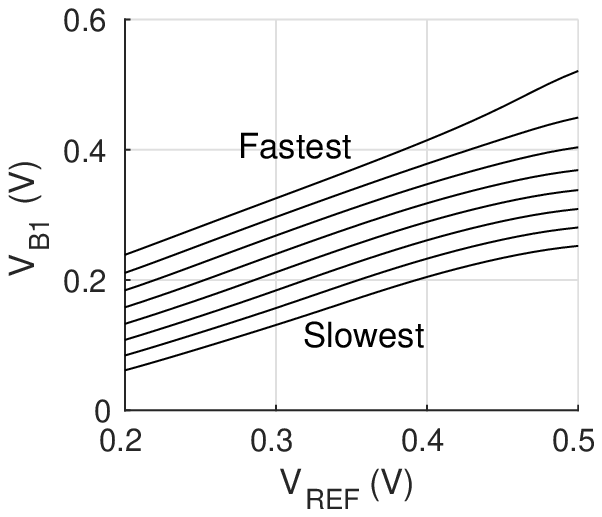}%
		}\\
		\subfloat[Output $V_{B2}$]{%
			\includegraphics[width=0.45\linewidth]{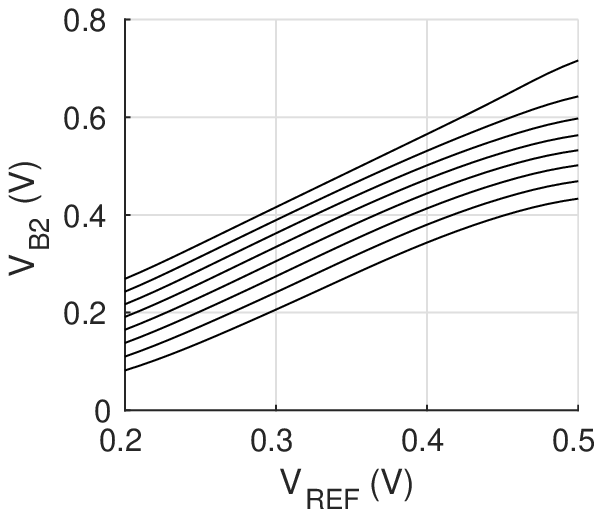}%
		}
		\subfloat[ $V_{B2}-V_{B1}$ ]{%
			\includegraphics[width=0.45\linewidth]{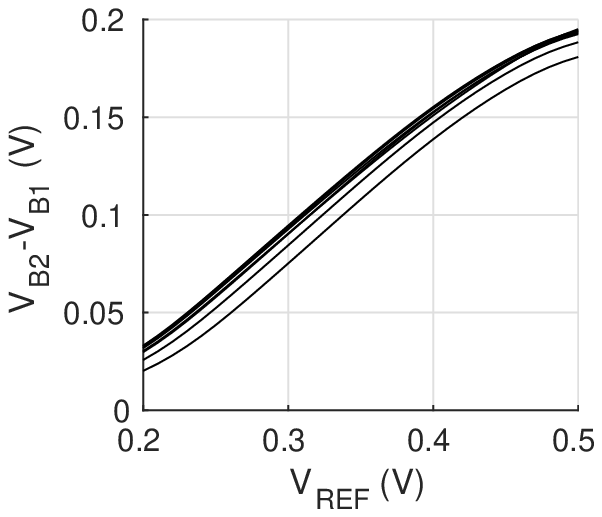}%
		}
		\caption{Biasing circuit outputs}
		\label{bias_plots}
	\end{figure}

 In the scaling branch, $M_{7-9}$'s widths are down-scaled by $2^{-n}$ w.r.t. the source cascode's width, which down-scale the current in the same proportion. $M_{12}$ is self-biased to accept the current and generates the primary bias $V_{B1}$. The gate voltage of the FET with the largest current exponent (or, the smallest delay exponent) is:
	\begin{equation}
	V_{B1,f} \approx V_{REF}
	\end{equation}
	Its drain-voltage is maintained at $100 mV$ using a fixed biased $M_{11}$. After down-scaling $M_{11}$'s size (by $2^{-i}$, $i$ being the exponent), its source voltage is maintained at a constant value. $M_{10}$, also a down-scaled transistor, is used to generate the secondary bias ($V_{B2}$). $M_{10}$'s width is adjusted using parametric analysis to keep its self-bias above $V_{B1}$ by $100mV$. An additional exponent-dependent scaling for the $M_{10}$ is needed, given by:
	\begin{equation}
	W_{M10,i}\approx{(1.3)^{-i}W_{M10,max}},
	\end{equation}
	
	to counter the lower turn-on voltages is required for the scaling branches. The response of the bias circuit is plotted in Fig. \ref{bias_plots}. As $V_{REF}$ varies, the bias current input to the multiplier's cascode is plotted in Fig. \ref{bias_plots}. $V_{B1}$, $V_{B2}$ and $V_{B2}-V_{B1}$ of down-scaled multiplier branches (upto 8 bits) is plotted in Fig. \ref{bias_plots}.
	
	\par To minimize distortion, it is essential that $M_7$'s corresponding to all exponents are applied the same drains-source voltage. Besides using a 3-level cascoding, the length of all transistors within the cascode is increased by $10\times$ over the minimum to minimize the CLM and short-channel effects that shoot-down the $r_{DS}$. The sizes of all transistors are summarized in Table \ref{tab_biasckt}.

	\begin{table}[t]
		\renewcommand{\arraystretch}{1.3}
		\caption{FET sizes for the Biasing Circuit}
		\centering
		\begin{tabular}{c c c}
			\toprule
			FET Label & Width ($/160 nm$) & Length ($/120nm$)\\
			\midrule
			$M_{1}$ & $1$ & $1$ \\
			$M_{2-3}$ & $2^n$ & $1$ \\
			$M_{4-6}$ & $10\times2^n$ & 10\\
			$M_{7-9}$ & $10\times2^i$ & $10$\\
			$M_{10}$ & ${2.6^i}$ & $10$\\
			$M_{11}$ & $2^i$ & $10$\\
			$M_{12}$ & 1 & 1\\
			\bottomrule
		\end{tabular}
		\label{tab_biasckt}
	\end{table}
	
	\subsection{Multiplier Simulation}
	Using the peripheral elements described in Sec. \ref{sec_backg}, the multiplier is simulated using transient simulators, for IBM $130nm$ technology.
	
	\subsubsection{Functionality test}
\begin{figure}[t]
		\centering
		\subfloat[Iso-$|W|$]{%
			\includegraphics[width=0.5\linewidth]{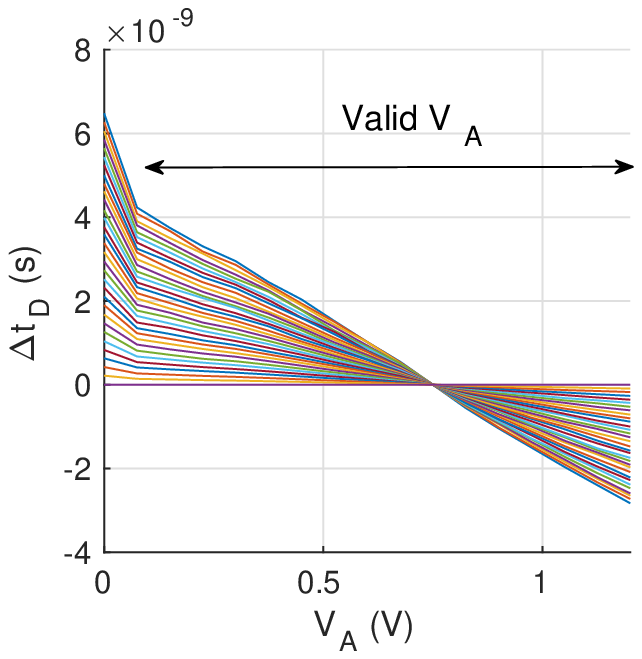}%
		}
		\subfloat[Iso-$V_A$]{%
			\includegraphics[width=0.5\linewidth]{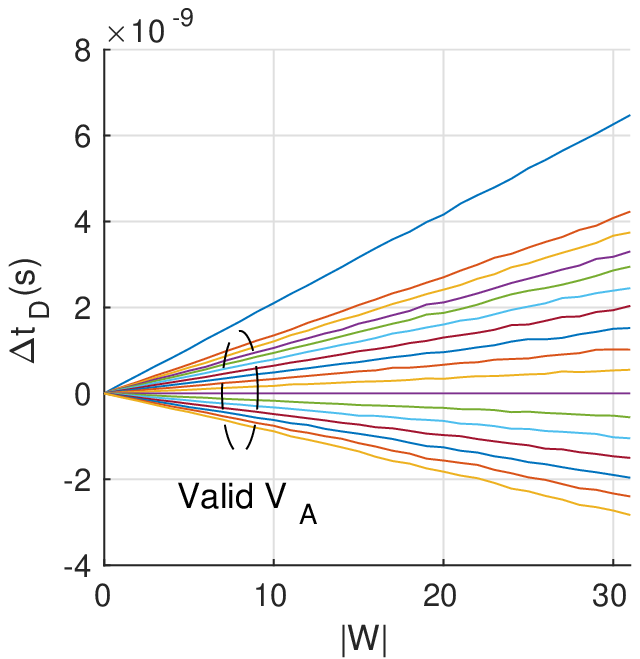}%
		}
		\caption{5-bit multiplier transfer characteristics}
		\label{mult_tc}
	\end{figure}

	A 5-bit multiplier, composed of delay-cells cascaded as described in Sec. \ref{sec_backg}, was simulated.
	Letting $V_{A0}=0.75V$, Fig. \ref{mult_tc}a plots the ref. delay of the multiplier as it varies with $V_A$, with weight ($W$) as a parameter. Conversely, ref. delay with $W$ as the independent variable and $V_A$ as parameter is plotted in Fig. \ref{mult_tc}b.
	Since the output ref. delay is distorted for $V_A<75mV$, the valid range of inputs for the multiplier is $75 mV$ to $1.2 V$.
	
	\subsubsection{Energy analysis and simulation results}
	Within a delay cell, the key components of energy are: 
	
	\begin{enumerate}
		
		\item \emph{$E_{C^*}$}: Energy used up in charging $C^*$ for each computation. It is given by:
		\begin{equation}
		E_{C^*}=C^*V_{DD}^2
		\end{equation}
		The actual value may vary due to parasitic capacitance and the dependence of $C^*$ on $V^*$
		\item \emph{$E_{TD}$}: Energy stored in node corresponding to $V_{RE}$ (\ref{dcs}), once $\Delta V^*$ crosses the threshold. It is given by: 
		\begin{equation}
		E_{TD}=C_{RE}V^2_{DD}
		\end{equation}
		where, $C_{RE}$ is the net capacitance at the node. Part of it comes from the thresholding-pFET $M_8$ ($E_{TD1}$) and other comes from latching-pFET, $M_{11}$ ($E_{TD2}$)
		\item \emph{$E_{PU}$}: Energy used in pull-up of the event-propagating wires of the delay cell ($OFE$ node in Fig. \ref{dcs})
		\item \emph{$E_{INV}$}: Energy used up in inverting the input falling-edge, to a rising edge ($IFE'$, input to $M_6$ in Fig. \ref{dcs})
		
	\end{enumerate}
	\par
	Next, value of these metrics is determined by simulating the 5-bit multiplier (schematic) for one cycle of computation, with arguments $V_A=1.2V$ and $\left|W\right|=31$. $E_{TD}$, $E_{PU}$ and $E_{INV}$ is determined during the computation-phase and $E_{C^*}$ and $E_{PU}$ are determined during pre-charge phase.
	The energy components and their simulated values are listed in Table \ref{tab_energy}. Comparing their sum with the simulated total, it is concluded that the listed components account for almost all the expended energy.
	
	\begin{table}[!t]
		\renewcommand{\arraystretch}{1.3}
		\caption{Delay-cell Energy Components}
		\centering
		\begin{tabular}{ c c c }
			\toprule
			Component & Energy/MAC (fJ) & Energy/MAC/bit (fJ)\\
			\midrule
			$E_{C^*}$ & 34.0 & 6.8\\
			$E_{TD1}$ & 5.6 & 1.1\\
			$E_{TD2}$ & 8.8 & 1.7\\
			$E_{PU}$ & 46.0 & 9.0\\
			$E_{INV}$ & 16.0 & 3.0\\
			\addlinespace
			Total & 110 & 22\\
			Total (sim.) & 116 & 23\\
			\bottomrule
		\end{tabular}
		\label{tab_energy}
	\end{table}

	\section{Discussion and Bench-marking}
	\label{sec_discuss}
	\par
\begin{figure}[t]
		\centering
		\includegraphics[scale=0.75]{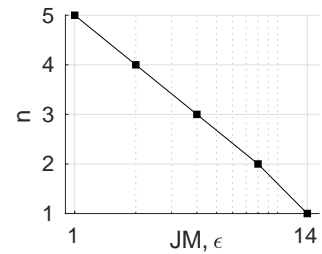}
		\caption{Number of bits vs. excess jitter margin}
		\label{jit_marg}
	\end{figure}

	In Sec. \ref{subsubsec_jitter_lim}, constraint on jitter was chosen such that the peak-jitter ($3\Delta t_{dN}$) from the slowest delay-cell, is less than the maximum referential delay of the fastest delay-cell. However, for certain inputs, the net output referential delay of a multiplier can be zero, which makes it impossible for for the noise to ever be smaller than the output signal. Thus, the chosen constraint is a practical as it grants the benefit of lower energy consumption by delay-based analog computing and simultaneously prevents excessive signal corruption by the noise. Depending on the signal-to-noise specification for an application, much tighter constraint on noise may be placed, which, in effect, reduces the maximum number of bits accomodable. Fig. \ref{jit_marg} plots the number of bits possible within a multiplier, as a function of excess jitter margin ($\epsilon$), where, $\epsilon$ is the ratio of maximum ref. delay of the fastest cell and peak-jitter. At $\epsilon=1$, the number of bits is the highest, and decreases to 1 at $\epsilon\approx14$.
	\begin{table*}[t]
		\renewcommand{\arraystretch}{1.3}
		\caption{Comparison with State-of-Art}
		\label{table_bm}
		\centering
		\begin{tabular}{| p{1.5cm} | p{1.5cm} | p{1.5cm} | p{1.5cm} | p{1.5cm} | p{1.5cm} | p{1.5cm} | }
			\hline
			& This work & Gopal et al. \cite{Gopal2018ACMOS} & Miyashita et al. \cite{Miyashita2014AnProcessing} & Sayal et al. \cite{Sayal201914.4Computing} & Lee et al. \cite{Lee2017AnalysisComputing} & Skrzyniarz et al. \cite{Skrzyniarz201624.3CMOS}\\
			\hline
			Domain & Time: delay & Time: delay & Time: Clocked-delay & Time: Pulse-width & Analog-charge & Analog-current\\
			\hline
			Demo. node & 130nm & 65nm & 65nm & 40nm & 40nm & 65nm\\
			\hline
			Input-width & Analog-5b & Analog-3b & 1b & 8b & Analog-3b & 2b/1b\\
			\hline
			Energy (fJ/MAC/bit) & 23 & 7 & 20 & - & 15 & 13\\
			\hline
			Latency & 1b: 1.2ns and 5b: 50ns & 250ps & 50ps & - & - & -\\
			\hline
			Negative weights & Yes & No & Yes & Yes & No & No\\
			\hline
			Linearity mechanism & Discharge-till-pinch-off & Body-gate biasing & Binary & Binary & NA & NA\\
			\hline
		\end{tabular}
		\label{tab_bm}
	\end{table*}
	\par
	Note that the multiplier's $E_{C^*}$, given in Table \ref{tab_energy}, is computed for the case when all weight bits are set to 1 (W=31). Otherwise, this components of energy depends on (1) the input weight and (2) number of computations being done by the MAC, per second. If the multiplier is used in a \emph{sense}, or, one-time-use mode, then, the listed $E_{C^*}$ is accurate, as all the charged-up energy leaks out eventually. Any new MAC cycle would require the same energy to charge-up $C^*$ from the point of no charge. However, for acceleration mode, where same weights are used with variable $V_A$, $C^*$ of the cells with $w=0$ never get an opportunity to discharge completely, since all falling-edges bypass the cell. Before it fully discharges, a new MAC cycle's pre-charge step would charge-up $C^*$ to $V_{DD}$ from intermediate voltage.
	
	\par
	In Table \ref{tab_bm}, two simulated performance metrics are compared with the state-of-art: (1) energy consumption per multiply-accumulate, reported above, and (2) multiplication latency (from absolute delay of the delay-cell). We also compare whether the multiplier allows negative weights and the maximum number bits accommodable for various mixed-signal MACs. From the table, it is seen that:
	\begin{enumerate}
		\item The delay cell consumes $23 fJ$ per bit of digital argument/input, which, is more than lowest-reported state-of-art energy consumption.  Our energy metric is at $130nm$, and the lowest state-of-art metric at $65nm$. Assuming that the energy scales by $L^2$, the scaled energy consumption approaches that of the state-of-art
		\item In \cite{Gopal2018ACMOS}, linearity of the delay-cell is based on back-body biasing, which, is theoretically non-linear. In the proposed cell, the output-input characteristics are linear, due to the linear voltage-initialization step
		\item Despite noise limitations, the number of bits that can be accommodated in the mixed-signal multiplier is higher than state-of-art. All reported mixed-signal MACs use an exponential scaling of transistor widths, as a way to convert digital signals to analog. In this work, we proposed a biasing circuit that exponentially scales the currents via gate-voltages and avoid area expensive width-scaling
	\end{enumerate}

	\section{Conclusion}
	\label{sec_concl}
	In this work, a linearly tunable delay-cell is proposed that realizes the analog input-dependent delay using three sequential sub-processes: (1) an input-dependant charge-up of $C^*$ (2) its steady discharge, via current $I^*$  (3) thresholding of its voltage.
	Each of the sub-processes is then analytically modeled, using which, constraints on the $C^*$ and $I^*$ for linearity are found.
	Jitter models, based on prior ones developed for CMOS inverter ring-oscillator, were modified and validated for the proposed cell.
	To form a multiplier, delay-cells with same analog input and $I^*$ scaled in the exponents of 2, must be cascaded to form a multiplier.
	Since $I^*$ is scaled using gates-source voltages, a biasing circuit that accept a ref. voltage and generates gate biases for delay cells corresponding to all exponents, is proposed and validated.
	From the constraints on $C^*$ for linearity and noise, the minimum $C^*$ was found to be around $2 fJ$ and maximum bits supportable to be five.
	Lastly we also identify key energy components of the multiplier, which sum up to be 20 fJ/MAC/bit for IBM's 130nm technology.
	
	\IEEEtriggeratref{17}
	
	
	\bibliographystyle{IEEEtran}
	\bibliography{main}

	
	

\end{document}